\documentclass[10pt,conference,final,a4paper]{IEEEtran}

\usepackage{cite}   
\usepackage{color}   
\usepackage[caption=false, font=footnotesize]{subfig}
\usepackage[pdftex]{graphicx}
\usepackage[latin1]{inputenc}
\usepackage{url}       
\usepackage{amsfonts}
\usepackage{amsmath}
\usepackage{times}
\usepackage{algorithm}
\usepackage[noend]{algpseudocode}

\usepackage{enumerate}
\usepackage{siunitx}
\usepackage{multirow}
\usepackage{tikz}

\makeatletter
\newcommand*{\algrule}[1][\algorithmicindent]{\makebox[#1][l]{\hspace*{.5em}\vrule height .75\baselineskip depth .35\baselineskip}}%

\newcount\ALG@printindent@tempcnta
\def\ALG@printindent{
    \ifnum \theALG@nested>0
        \ifx\ALG@text\ALG@x@notext
            \addvspace{-0.5pt}
        \else
            \unskip
            \ALG@printindent@tempcnta=1
            \loop
                \algrule[\csname ALG@ind@\the\ALG@printindent@tempcnta\endcsname]
                \advance \ALG@printindent@tempcnta 1
            \ifnum \ALG@printindent@tempcnta<\numexpr\theALG@nested+1\relax
            \repeat
        \fi
    \fi
    }
\usepackage{etoolbox}
\patchcmd{\ALG@doentity}{\noindent\hskip\ALG@tlm}{\ALG@printindent}{}{\errmessage{failed to patch}}
\makeatother

\newlength{\oldtextfloatsep}
\begin{document}
\title{Optimized Network-coded Scalable Video Multicasting over eMBMS Networks}

\author{\IEEEauthorblockN{Andrea Tassi\IEEEauthorrefmark{1}, Ioannis Chatzigeorgiou\IEEEauthorrefmark{1}, Dejan Vukobratovi\'c\IEEEauthorrefmark{3} and Andrew L. Jones\IEEEauthorrefmark{1}}
\IEEEauthorblockA{\IEEEauthorrefmark{1}School of Computing and Communications, Lancaster University, United Kingdom}
\IEEEauthorblockA{\IEEEauthorrefmark{3}Department of Power, Electronics and Communication Engineering, University of Novi Sad, Serbia}}

\maketitle

\begin{abstract}
Delivery of multicast video services over fourth generation (4G) networks such as 3GPP Long Term Evolution-Advanced (LTE-A) is gaining momentum. In this paper, we address the issue of efficiently multicasting layered video services by defining a novel resource allocation framework that aims to maximize the service coverage whilst keeping the radio resource footprint low. A key point in the proposed system mode is that the reliability of multicast video services is ensured by means of an Unequal Error Protection implementation of the Network Coding (UEP-NC) scheme. In addition, both the communication parameters and the UEP-NC scheme are jointly optimized by the proposed resource allocation framework. Numerical results show that the proposed allocation framework can significantly increase the service coverage when compared to a conventional Multi-rate Transmission (MrT) strategy.
\end{abstract}


\section{Introduction}\label{sec:intro}
Over the last few years, the technological evolution of communication devices has fuelled a surge in demand for new multimedia services over fourth generation (4G) and \mbox{next-generation} cellular networks. To give an example, up to $67$\% of the global mobile Internet traffic will be represented by video content by 2018~\cite{CVI}. Various standards for video coding and compression have been proposed to enable layered video streaming. Among them, the H.264 Scalable Video Coding (H.264/SVC) standard is gaining popularity~\cite{6025326}. 

The key idea behind H.264/SVC compression is the generation of a layered video stream, which consists of one \emph{base layer} and multiple \emph{enhancement layers}. The base layer ensures a basic reconstruction quality, which progressively improves with the number of recovered enhancement layers. Hence, a scalable video service can be transmitted to multiple network users and potentially be decoded by them at different quality levels, depending on the radio propagation conditions.

In this paper, we consider a 3GPP Long Term Evolution-Advanced (\mbox{LTE-A}) communication network multicasting H.264/SVC video services. Since its first release, the \mbox{LTE-A} standard manages Point-to-Multipoint (PtM) services by means of the evolved Multimedia Broadcast and Multicast Service (eMBMS) framework~\cite{dahlman20114g}. A PtM service can be delivered by using the Single-Cell (SC-) or the Single Frequency Network (SFN-) eMBMS modes. In the SC-eMBMS mode, the service is delivered by each base station independently of the others. On the other hand, the SFN-eMBMS mode allows multiple neighbouring base stations (forming the SFN) to deliver the same PtM service in a synchronous fashion. Hence, base stations of the same SFN do not interfere with each other.

3GPP addresses the reliability issues of PtM communications by proposing the adoption of Application Level-Forward Error Correction (AL-FEC) schemes based on Raptor codes~\cite{6353684}. However, Magli~\textit{et al.}~\cite{6416071} noted that this kind of sparse codes, as well as LT codes, are usually designed to be applied over large source messages. As a consequence, the use of Raptor or traditional LT codes can lead to a \mbox{non-negligible} communication delay. As noted in~\cite{6416071}, this issue can be mitigated by using Random Linear Network Coding (RLNC) strategies, applied over short source messages. 

This paper draws inspiration from~\cite{6353397} and~\cite{Tassi}, which utilize the RLNC principle to protect the reliability of Point-to-Point (PtP) and PtM communications, respectively. Unlike~\cite{6353397} and~\cite{Tassi}, this paper takes into account the fact that the layers of a video stream have different importance levels and hence we adopt the Unequal Error Protection (UEP) implementation of RLNC~\cite{6168183}, which we shall refer to as UEP-NC. UEP-NC allows the transmitter to adjust the error protection capability of the code according to the importance level of the transmitted video layer.

In addition to tackling reliability issues, the definition of efficient resource allocation models suitable for delivering PtM services is also a challenging task. Among candidate resource allocation models, the family of Multi-rate Transmission (MrT) strategies is the most fitting for exploiting the layered nature of a service~\cite{6148193}. MrT strategies attempt to split users into subgroups. The transmission parameters of each video layer are then optimized by considering the propagation conditions experienced in each user subgroup. In this way, all subgroups receive the same video stream but each subgroup is able to recover a different set of video layers and, thus, experience a different level of video quality~\cite{5452675,4917957}.

A MrT resource allocation model can be designed to achieve a variety of objectives. For instance,~\cite{5452675,4917957} maximize the total service level experienced by all the users. On the other hand,~\cite{Tassi} concentrates on minimizing the amount of radio resources needed to broadcast a video service. In contrast to~\cite{5452675,4917957,Tassi}, this paper interprets the number of recovered video layers as the \emph{system profit}, while the amount of radio resources required to deliver the PtM service is considered the \emph{system cost}. According to a fundamental economics principle, we propose a novel MrT-based allocation model, which we call UEP Resource Allocation Model (UEP-RAM). Our model maximises the system profit-cost ratio and ensures that at least a desired fraction of users achieve a predetermined service level. A key aspect in the proposed UEP-RAM is that system and UEP-NC parameters are jointly optimized.

The rest of the paper is organized as follows. Section~\ref{sec:sysMod} presents the theoretical framework for the assessment of the service level achieved by each user. The integration of the UEP-NC principle into the LTE-A protocol stack is also presented. Section~\ref{sec:RA} describes the proposed UEP-RAM and provides a novel heuristic strategy to efficiently derive a good quality solution. Results are discussed in Section~\ref{sec:Res} while the main findings of the paper are summarized in Section~\ref{sec:Concl}.

\section{System Model}\label{sec:sysMod}
We consider an eMBMS network that delivers the same \emph{target scalable video stream} to a multicast group of $U$ users. If the eMBMS network consists of a single base station, the video service is transmitted according to the SC-eMBMS mode. On the other hand, if the eMBMS network is formed by two or more spatially contiguous base stations, they will form a SFN~\cite{dahlman20114g}.

\subsection{Unequal Error Protection Random Linear Network Coding}\label{subsec:UEP-NC}
\begin{figure}[tbd]
\centering
\includegraphics[width=0.8\columnwidth]{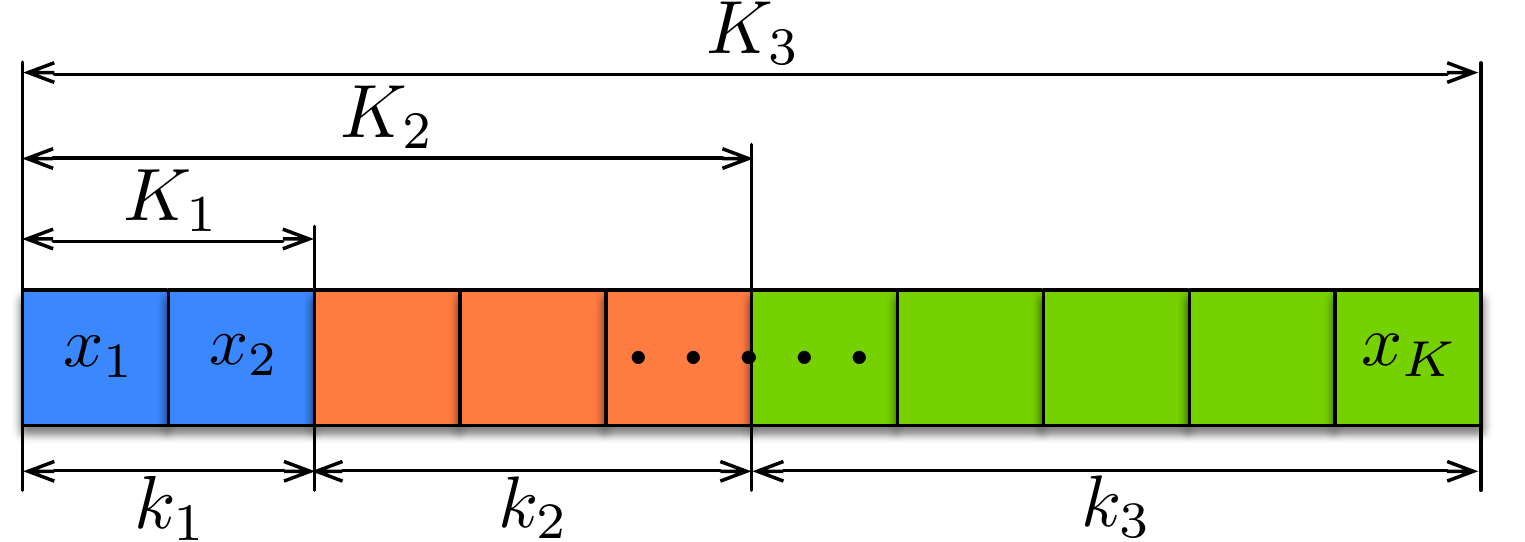}
\caption{Example of layered source message $\mathbf{x}$ comprised of $L = 3$ layers.}
\label{fig.SourceMessage}
\vspace{-3mm}
\end{figure}

We denote by $\mathbf{x} = \{x_1, \ldots, x_K\}$ a \emph{scalable} source message composed of $K$ elements. As shown in Fig.~\ref{fig.SourceMessage}, the elements of $\mathbf{x}$ can be grouped into $L$ \emph{layers}. Each layer signifies a different importance level and consists of a fixed number of elements. For example, the $\ell$-th layer consists of $k_\ell$ elements. We assume that layers are arranged in descending order of importance, that is, the first layer -- which comprises the first elements of the message -- is the most important layer, while the $L$-th layer is the least important layer. We define the user Quality-of-Service (QoS) level as the number of \emph{consecutive} layers that a user can recover, starting from the first layer.  

In an effort to improve reliability, each source message is transmitted to users according to the RLNC principle~\cite{Medard}. A typical RLNC PtM scheme generates a stream of \mbox{$N\geq K$} coded elements $\mathbf{y} = \{y_1, \ldots, y_N\}$ obtained by linearly combining elements of $\mathbf{x}$. Hence, coded element $y_j$ is defined as $y_j = \sum_{i = 1}^K g_{j,i}\,x_i$, where each \emph{coding coefficient} $g_{j,i}$ is selected uniformly at random over a finite field $\mathrm{GF}(q)$ of size $q$. A user can recover the source message $\mathbf{x}$ as soon as it collects $K$ linearly independent coded elements~\cite{Medard}.

A key issue of conventional RLNC-based schemes is that a user is not able to recover all the $K$ source elements if it has not collected at least $K$ linearly independent coded elements. To address this issue, we adopt the UEP implementation of the RLNC~\cite{6168183}, which we shall refer to as UEP-NC. In this case, instead of generating coded packets by linearly combining source packets over the whole message $\mathbf{x}$, the RLNC process takes place over a nested structure of $L$ \emph{expanding windows}. As depicted in Fig.~\ref{fig.SourceMessage}, the $\ell$-th expanding window $\mathbf{x}_{1:\ell}$ consists of all the source elements belonging to the first $\ell$ layers. In other words, $\mathbf{x}_{1:\ell}$ is formed by the first $K_\ell = \sum_{i = 1}^{\ell}k_i$ elements of the source message $\mathbf{x}$.

In this paper, we map coded elements associated with an expanding window onto Packet Data Units (PDUs). The PDU description and its role in the LTE-A protocol stack will be clarified in Section~\ref{subsec:SVC}. For the sake of simplicity, we assume that coded elements associated with different expanding windows are not mixed within the same PDU. The selection of the coding coefficients for the generation of each coded element is governed by a Random Number Generator (RNG). As explained in~\cite{6353397}, the RNG seed associated with the first coded element in a PDU is transmitted along with standard LTE-A signalling information. The RNG seeds associated with the remaining coded elements in the PDU are then incrementally computed from that associated with the first coded element.

Let us focus on one of the $N_\ell$ PDUs associated with the $\ell$-th expanding window $\mathbf{x}_{1:\ell}$. This PDU can hold $n_\ell$ coded elements\footnote{As a consequence, the total number of transmitted coded elements associated with the $\ell$-th expanding window is $N_\ell\: n_\ell$.}, which are transmitted over a broadcast erasure channel using a Modulation and Coding Scheme (MCS) with index $m_\ell$. The probability that user $u$ will fail to recover the PDU is denoted by $p_{u,\ell}$. Note that the larger the value of $m_\ell$ is, the higher the modulation order or the lower the error-correcting capability of the corresponding MCS is. We can infer that the value of $m_\ell$ has an impact on both $p_{u,\ell}$ and $n_\ell$; if $m_\ell$ increases, then $n_\ell$ and $p_{u,\ell}$ are also likely to increase.

Based on our adopted model, if $N_\ell$ PDUs associated with expanding window $\mathbf{x}_{1:\ell}$ are transmitted, $r_\ell \leq N_\ell$ of them will be successfully received by user $u$. We initially refer to the case where source elements of expanding window $\mathbf{x}_{1:\ell}$ can be retrieved from PDUs related to the first $\ell$ expanding windows. For convenience, we introduce sets $\mathbf{N}_{1:\ell} = \{N_1, \ldots, N_\ell\}$ and $\mathbf{r}_{1:\ell} = \{r_1, \ldots, r_\ell\}$ to enumerate all transmitted and received PDUs, respectively, which are associated with each of the first $\ell$ expanding windows. The probability that user $u$ achieves the QoS level $\ell$ can be quantified by the probability $\textrm{P}_u(\mathbf{N}_{1:\ell})$ of recovering the source elements of the $\ell$-th expanding window, expressed as
\begin{equation}
\textrm{P}_u(\mathbf{N}_{1:\ell}) \!=\!\! \sum_{r_1 = 0}^{N_1} \!\! \cdot\cdot\!\! \sum_{r_\ell = 0}^{N_\ell} \overbrace{\prod_{i = 1}^{\ell} \binom{N_i}{r_i} \, (1\!-\!p_{u,i})^{r_i} \, p_{u,i}^{N_i - r_i}}^{\textrm{R}_u(\mathbf{r}_{1:\ell})}  g(\mathbf{r}_{1:\ell})\label{eq:P2}
\end{equation}
where $\textrm{R}_u(\mathbf{r}_{1:\ell})$ represents the probability that user $u$ successfully receives $r_i$ out of $N_i$ coded elements, for $i = 1, \ldots, \ell$. Function $g(\mathbf{r}_{1:\ell})$ is the probability of recovering the $\ell$-th expanding window, given that user $u$ has successfully collected the set of PDUs described by $\mathbf{r}_{1:\ell}$.

To the best of our knowledge, a closed form expression for $g(\mathbf{r}_{1:\ell})$ is not known. However, we know that if the field size $q$ tends to infinity, the probability of receiving a coded element, which is linearly independent of previously received coded elements, approaches one. Following the rationale of~\cite{jsacTassi} and considering large values of $q$, we set $g(\mathbf{r}_{1:\ell})=1$ if the number $r_\ell\:n_\ell$ of successfully received coded elements associated with the $\ell$-th expanding window is equal to or greater than a threshold value $r_{\text{min}, \ell}$; otherwise, we set $g(\mathbf{r}_{1:\ell})=0$. We can thus write
\begin{equation}
g(\mathbf{r}_{1:\ell}) \simeq I\left(r_\ell \, n_\ell \geq r_{\text{min},\ell}\right)\label{eq:g}
\end{equation}
where $I(s)$ is the indication function, i.e., $I(s)=1$ if statement $s$ is true, otherwise $I(s)=0$. Using the same reasoning as in~\cite{jsacTassi}, we can express $r_{\text{min}, \ell}$ as the following recursion
\begin{equation}
r_{\text{min},\ell} = 
K_\ell - K_{\ell-1} + \max\left(r_{\text{min},\ell-1}\:-\:r_\ell\: n_{\ell-1},\:\:0\right)
\end{equation}
for an initial value of $r_{\text{min},1}=K_1$.

\begin{figure}[t]
\centering
\hspace*{-1.5mm}\subfloat[$n_\ell = 2$ for $\ell = 1, \ldots, L$]{\label{fig.val2}
\includegraphics[width=0.505\columnwidth]{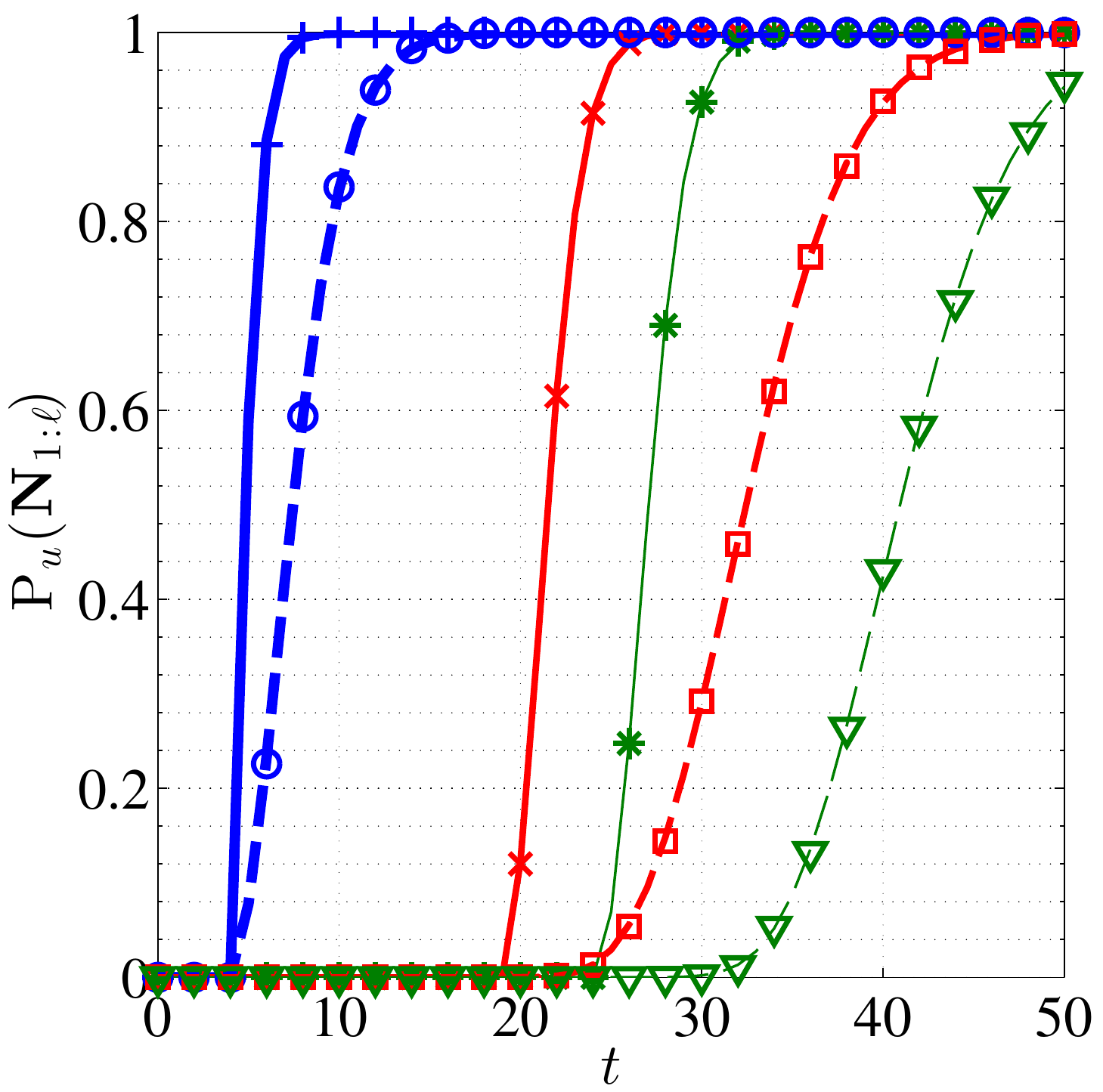}
}
\hspace*{-2.5mm}\subfloat[$n_\ell = 5$ for $\ell = 1, \ldots, L$]{\label{fig.val28}
\includegraphics[width=0.505\columnwidth]{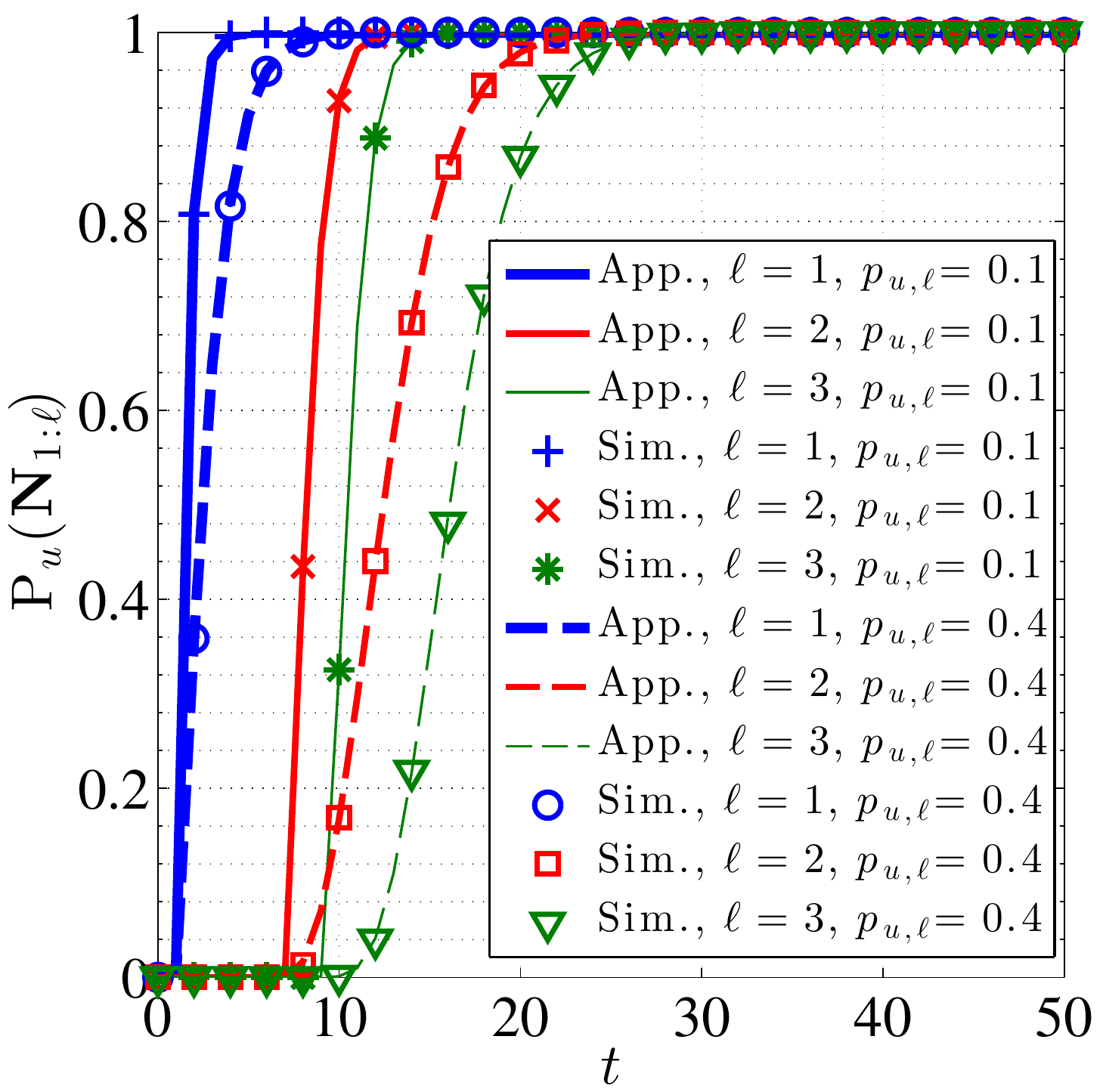}
}
\caption{Performance comparison between the approximated (``App.'') and simulated (``Sim.'') version of $\textrm{P}_u(\mathbf{N}_{1:\ell})$.}
\label{fig.val}
\vspace{-5.9mm}
\end{figure}

In this paper, we invoke approximation~\eqref{eq:g} for the calculation of~\eqref{eq:P2}. To validate the accuracy of the proposed approximation, we compare the probability values derived from~\eqref{eq:P2} with those obtained by simulation. Fig.~\ref{fig.val} shows both the approximated and simulated probability values of $\textrm{P}_u(\mathbf{N}_{1:\ell})$ in the case of a three-layer source message with $K_1 = 10$, $K_2 = 50$ and $K_3 = 100$, when the finite field has $q = 2^8$ elements. We assume that each PDU can hold the same number of coded elements ($n_\ell = 2$ or $5$) and is affected by the same erasure probability ($p_{u,\ell} = 0.1$ or $0.4$). Probabilities have been plotted as a function of $t$, where $N_\ell = t$, for $\ell = 1, 2, 3$. We observe that the performance gap between simulation and approximated calculations is smaller than $7 \cdot 10^{-3}$, which confirms that the proposed approximation is accurate for $q \geq 2^8$. In particular, in the rest of the paper, we will refer to $q = 2^8$.

\subsection{Layered Video Multicasting over eMBMS Networks}\label{subsec:SVC}

Let us first consider a network operating in SC-eMBMS mode within which a single base station delivers a H.264/SVC video stream. The stream consists of $L$ layers $\{v_1, \ldots, v_L\}$, where $v_1$ represents the base video layer, and $\{v_2, \ldots, v_L\}$ represent the $L-1$ enhancement video layers. We remark that the user QoS improves with the number of recoverable \emph{consecutive video layers}, starting from $v_1$.

In our system model, each data stream associated with a video layer traverses the Packet Data Conversion Protocol (PDCP), the Radio Link Control (RLC) and the MAC layers of the LTE-A protocol stack of the base station. In this paper, similarly to the protocol stack presented in~\cite{6353397}, we make use of a modified implementation of the standard MAC layer, which allows the transmission of the video stream by means of the UEP-NC principle as explained in Section~\ref{subsec:UEP-NC}.

The output of an H.264 video encoder can be modelled as a stream of Group of Pictures (GoPs), each GoP has a duration of $d_{\text{GoP}}$ seconds. Since the video decoding process takes place on a GoP-by-GoP basis~\cite{6025326}, a GoP represents a layered source message. According to the layered model presented in Section~\ref{subsec:UEP-NC}, the $\ell$-th layer of a source message consists of $k_\ell$ elements. To fit our model, the data stream associated with the $\ell$-th layer of a GoP is segmented by the MAC layer into $k_\ell$ source elements. If $H$ is the number of bits comprising each source element and $b_\ell$ is the bit-rate of the $\ell$-th layer of the GoP, the number of source elements forming the $\ell$-th layer is $k_\ell = \left\lceil(b_\ell\:d_{\text{GoP}}) / H\right\rceil$.

\begin{figure}[t]
\vspace{-0.3mm}
\centering
\hspace*{-2mm}\subfloat[Considered LTE-A stack]{\label{fig.SysFrameA}
\includegraphics[width=0.5\columnwidth]{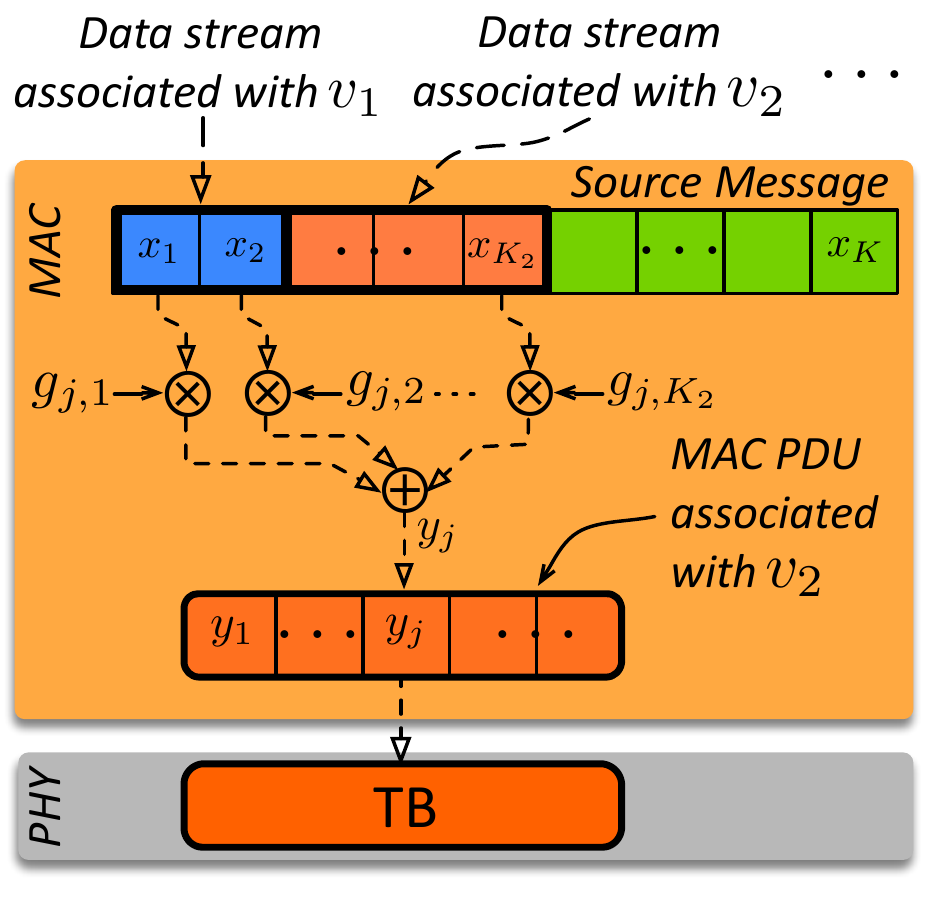}
}
\hspace*{-1mm}\subfloat[Radio frame model]{\label{fig.SysFrameB}
\includegraphics[width=0.51\columnwidth]{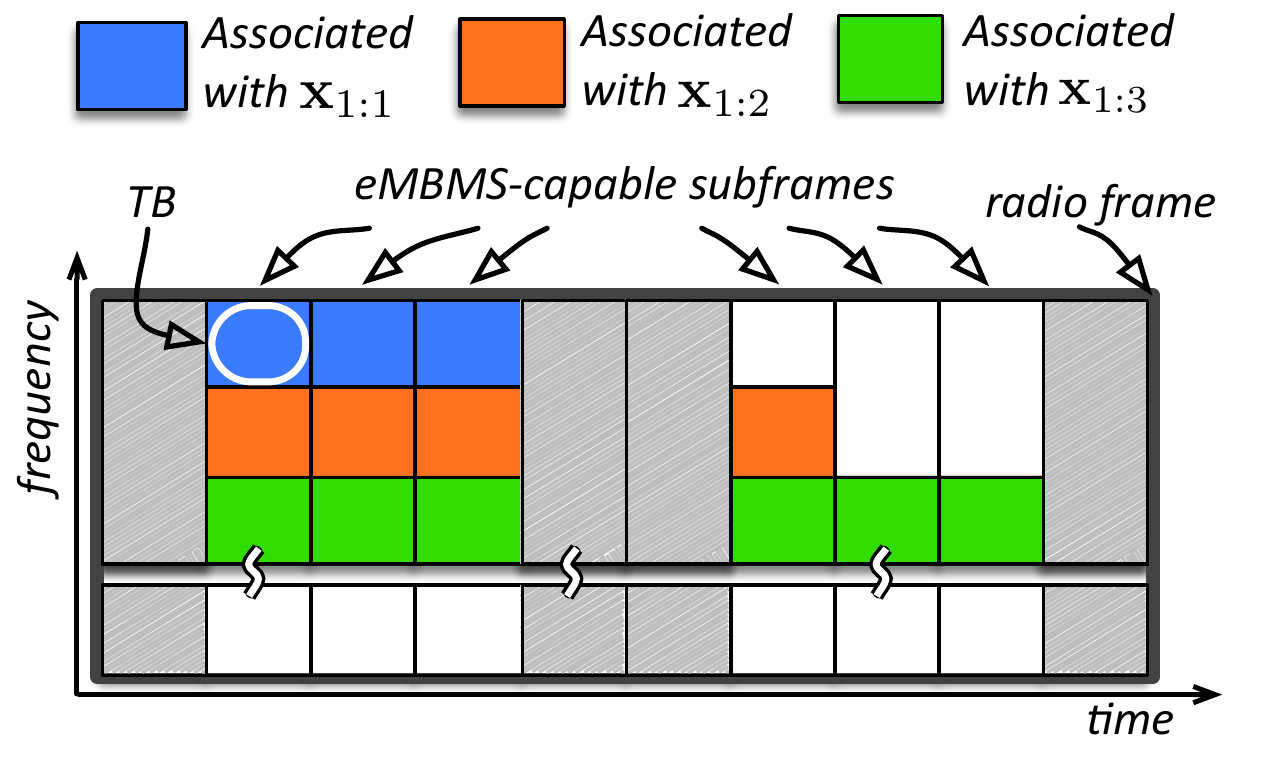}
}
\vspace{1.7mm}
\caption{Generation of a MAC PDU associated with $\mathbf{x}_{1:2}$ (left side) and TB allocation on an LTE-A radio frame for $L = 3$ (right side).}
\vspace{-5.4mm}
\label{fig.SysFrame}
\end{figure}

As shown in Fig.~\ref{fig.SysFrameA}, the modified MAC layer generates coded elements associated with the $\ell$-th expanding window. All coded elements are then mapped onto one or more MAC PDUs, each of which can hold $n_\ell$ coded packets. The sequence of $N_\ell$ MAC PDUs is then forwarded to the physical layer. The LTE-A standard imposes that a MAC PDU is mapped onto exactly one physical layer Transport Block (TB). For this reason, we use the terms PDU and TB interchangeably. 

Fig.~\ref{fig.SysFrameB} depicts the frequency-time structure of an \mbox{LTE-A} radio frame. Every radio frame comprises $10$ subframes, each of which has a time duration of one Transmission Time Interval (TTI), which is equal to $d_{\text{TTI}}= 1 \, \mathrm{ms}$. Since at most 6 subframes can convey eMBMS data traffic~\cite{dahlman20114g}, we use $f_{\text{eMBMS}} = 0.6$ to refer to the maximum fraction of \mbox{eMBMS-capable} subframes per radio frame. Notice that a TB is a fraction of the radio frame. It has a \emph{fixed} time duration of one TTI, and consists of $N_{\text{RBP}}$ Resource Block Pairs\footnote{A resource block pair can be modelled as a frequency-time structure spanning a bandwidth of $180 \, \mathrm{kHz}$ and a transmission time interval of 1 TTI.} (RBPs)~\cite{dahlman20114g}, not shown in Fig.~\ref{fig.SysFrameB}. The number $n_\ell$ of coded elements that a TB can hold depends on the adopted MCS, characterized by index $m_\ell$, the value of $N_{\text{RBP}}$ and the element size $H$. Table~\ref{tab.params} reports the values of the ratio $n_\ell/N_{\text{RBP}}$ for $H = 2$ KB and different MCS indexes. The reported values have been selected in order to fit the actual bit capacity of a TB~\cite{6600846}. 

Consider Fig.~\ref{fig.SysFrameB}, we assume that at most one TB associated with the same expanding window  can be transmitted during the same subframe. The remaining part of the radio frame that has not been reserved for the delivery of the layered video service under consideration (marked in white) can be used by other LTE-A eMBMS services.

We now shift our focus from the SC-eMBMS to the \mbox{SFN-eMBMS} mode. In this case, base stations belonging to the same SFN are connected to the Multicell Coordination Entity (MCE), which is in charge of synchronizing the transmissions of all base stations~\cite{dahlman20114g}. The LTE-A standard imposes that all SFN base stations multicast the same video stream using the same physical signals. To this end, we assume that all RNGs, used by the SFN base stations during the network encoding process, are kept synchronized by the MCE. Consequently, all SFN base stations generate the same set of coded elements.

In both SC- and SFN-eMBMS modes, we assume that each user can provide Channel Quality Indicator (CQI) feedback on the downlink channel conditions. CQI feedback is sent to the only base station in SC-eMBMS or the base stations that is closest to the user in SFN-eMBMS. In the latter case, the reported CQI is forwarded to the MCE. In the CQI feedback, each user specifies the index of the MCS that ensures the maximum transmission rate and a TB error probability or, equivalently, a MAC PDU erasure probability not greater than $\Hat{p} = 0.1$~\cite{dahlman20114g}.

\section{Scalable Video Multicasting Optimization}\label{sec:RA}
In this section, we propose UEP-RAM to maximize the QoS level experienced by each user during the transmission of scalable video streams, while keeping the number of TB transmissions low. The constraint set of the proposed model ensures that a predetermined fraction of users shall achieve the desired QoS level with at least a given probability. In the case of SC-eMBMS, we assume that the proposed resource allocation strategy is performed at the MAC layer along with the scheduling operations defined by the standard. In the case of SFN-eMBMS, the MCE is in charge of scheduling services within the SFN. For this reason, we assume that UEP-RAM is also handled by the MCE.

We denote by $m^{(u)} \in [1, 15]$ the MCS index reported in the CQI feedback of a user $u \in [1, U]$. Upon receiving the $m^{(u)}$ values for any value of $u$, base stations employ the same MCS with index $m_\ell$ to broadcast TBs associated with the $\ell$-th expanding window. If \mbox{$m_\ell \leq m^{(u)}$} for a user $u$, the TB error probability $p_{u,\ell}$ experienced by that user will not be greater than $\Hat{p}$. However, base stations cannot compute $\textrm{P}_u(\mathbf{N}_{1:\ell})$ using \eqref{eq:P2} as they are not aware of the exact value of $p_{u,\ell}$. To alleviate this problem, our proposed optimization strategy sets $p_{u,\ell} = \Hat{p}$ if $m_\ell \leq m^{(u)}$, otherwise $p_{u,\ell} = 1$.

Let $\delta_{u,\ell}$ be an indication variable associated with user $u$ and QoS level $\ell$, so that $\delta_{u,\ell}\!=\!1$ if the first $\ell$ service layers are recovered by user $u$ with a probability of at least $\Hat{Q}$; otherwise, $\delta_{u,\ell} = 0$. We understand that user $u$ will recover the first $\ell$ service layers if it successfully recovers the $\ell$-th expanding window or \textit{any} of the expanding windows having index greater than $\ell$. Thus, the indication variable $\delta_{u,\ell}$ can be expressed as
\begin{equation}
\delta_{u,\ell} = I\left(\bigvee_{i = \ell}^{L} \textrm{P}_u(\mathbf{N}_{1:i}) \geq \Hat{Q}\right)\label{eq.delta}
\end{equation}where $\vee$ denotes the logical OR operator.

We define the \emph{profit} made by the system as the number of video layers that any user can recover with a given probability. Using \eqref{eq.delta}, we can express the profit as $\sum_{u = 1}^U\sum_{\ell= 1}^L \delta_{u,\ell}$. On the other hand, we represent the \emph{cost} incurred by the system as the number of required TBs for the delivery of a GoP, which is captured by $\sum_{\ell = 1}^L N_\ell$. In order to optimize the profit while keeping the cost low, the proposed UEP-RAM aims at maximizing the \emph{profit-cost ratio} as follows
\begin{align}
	\text{(UEP-RAM)} &  \quad  \mathop{\max_{m_1, \ldots, m_L}}_{N_1, \ldots, N_L} \,\,  \sum_{u = 1}^U\sum_{\ell= 1}^L \delta_{u,\ell} \Big/ \sum_{\ell = 1}^L N_\ell \label{UEP.of}\\
   \text{subject to} &   \quad \sum_{u = 1}^U \delta_{u,\ell} \geq U \, \Hat{t}_\ell \quad\quad\quad\,\text{$\ell= 1, \ldots, L$} &\label{UEP.c1}\\
                     &   \quad 0 \leq N_\ell \leq \Hat{N}_\ell \quad\quad\quad\quad\text{$\ell= 1, \ldots, L$.}&\label{UEP.c2}
\end{align}

Objective~\eqref{UEP.of} represents the system profit-cost ratio. Constraint~\eqref{UEP.c1} ensures that the fraction of users recovering the first $\ell$ video layers with a probability of at least $\Hat{Q}$ shall not be smaller than $\Hat{t}_\ell$. Constraint~\eqref{UEP.c2} imposes that the transmission of data associated with the $\ell$-th expanding window shall not require more than $\Hat{N}_\ell$ TBs. Of course, the transmission of a GoP should not exceed the GoP duration $d_{\text{GoP}}$. For this reason, the values of $\Hat{N}_1, \ldots, \Hat{N}_L$ have to be such that $\max(\Hat{N}_1, \ldots, \Hat{N}_L) \leq \left\lfloor f_{\text{eMBMS}} \cdot d_{\text{GoP}}\,/\,d_{\text{TTI}} \right\rfloor$.

Unfortunately, the UEP-RAM is a computationally complex optimization problem because of the coupling constraints among optimization variables introduced by~\eqref{UEP.c1}. For this reason, we propose a novel heuristic strategy, summarized in Procedure~\ref{Alg.P1}, which can be used to find a good quality solution of the UEP-RAM in a finite number of steps.

\setlength{\textfloatsep}{5pt}
\begin{algorithm}[tbd]
\floatname{algorithm}{Procedure}
\caption{Heuristic UEP-RAM}
\label{Alg.P1}
\begin{algorithmic}[1]
\scriptsize
\State $s \gets L-1$\label{sInit}
\While {$s \geq 0$}\label{wStart}
	\State $N_\ell \gets 0$ and $m_\ell \gets 0$, for any $\ell \in [1, \ldots, L]$
	\State $t^\prime_{s+1} \gets \Hat{t}_1$, and $t^\prime_i \gets \Hat{t}_i$, for any $i \in [s+2, \ldots, L]$
	\State $m_\ell \gets \text{solution of S1-$\ell$ problem, for any $\ell \in [s+1, \ldots, L]$}$\label{s1Sol}
	\For{$\ell \gets s+1, \ldots, L$}\label{s2SolStart}
		\If{S2-$\ell$ can be solved} $N_\ell \gets \text{solution of S2-$\ell$}$
		\EndIf
	\EndFor\label{s2SolEnd}
	\vspace*{-0.04cm}\If{constraint~\eqref{UEP.c1} holds}\label{ifPreRef}
				\hspace*{3.3cm}\State $m_{\textrm{INT},\ell} \gets m_\ell$ and $N_{\textrm{INT},\ell} \gets N_\ell$, for any $\ell \in [1, \ldots, L]$\label{refstepStart}
			\For{$\ell \gets L, \ldots, s+2$}
			\If{$N_{\ell-1} > 0$ \textbf{and} $N_\ell > 0$}
				\State $m_\ell \gets m_{\ell-1}$, and $N_\ell \gets 0$
				\If{S2-$\ell$ can be solved}
					\State $N_\ell \gets \text{solution of S2-$\ell$}$, and $m_{\ell-1} \gets 0$
				\ElsIf{$N_\ell > N_{\textrm{INT},\ell}$}
					\State $m_\ell \gets m_{\textrm{INT},\ell}$, and $N_\ell \gets N_{\textrm{INT},\ell}$
				\EndIf			
			\EndIf
		\EndFor
		\vspace*{-0.04cm}\If {$\sum_{\ell = 1}^L N_{\textrm{INT},\ell} < \sum_{\ell = 1}^L N_\ell$}
			\State{\Return $m_{\textrm{INT},1}, \ldots, m_{\textrm{INT},L}$ and $N_{\textrm{INT},1}, \ldots, N_{\textrm{INT},L}$}
		\Else
\vspace*{-0.05cm}			\State\Return $m_1, \ldots, m_L$ and $N_1, \ldots, N_L$
	    \EndIf\label{refstepEnd}
\vspace*{-0.05cm}	\Else $\,\,\,s \gets s - 1$\label{sDecr}
	\EndIf
\EndWhile\label{wEnd}
\vspace*{-0.05cm}\State{\Return \textit{no solution found.}}
\end{algorithmic}
\end{algorithm} 

Consider constraint~\eqref{UEP.c2} which allows $N_\ell$ to be equal to $0$. In that case, coded elements associated with $\mathbf{x}_{1:\ell}$ will be neither generated nor transmitted. Let us assume that the first $s$ expanding windows are not delivered, where $0 \leq s \leq L-1$. The desired service coverage can only be offered if (i) the first $(s+1)$ service layers are recovered with a probability of at least $\Hat{Q}$ by the user fraction $t^\prime_{s+1} = \Hat{t}_1$, and (ii) the remaining layers are recovered by the user fractions $t^\prime_{i} = \Hat{t}_i$, where $i = (s+2), \ldots, L$.

The rationale for the proposed heuristic can be summarized as follows. The value of $s$ is initially set to $L-1$ \mbox{(line~\ref{sInit} of} Procedure~\ref{Alg.P1}). The while-loop (lines~\ref{wStart}-\ref{wEnd}) iteratively tries to find a feasible solution to the UEP-RAM. However, if constraints~\eqref{UEP.c1} and~\eqref{UEP.c2} are not met, then the value of $s$ is decreased (line~\ref{sDecr}). In particular, for any $\ell \in [s+1, L]$, the following problem is solved (line~\ref{s1Sol})
\begin{equation}
\text{(S1-$\ell$)} \,\, \displaystyle\mathop{\arg\max}_{m_\ell \in [1, \ldots, 15]} \Big\{\sum_{u = 1}^U I\left( m_\ell \leq m^{(u)}\right) \geq U \, t^\prime_\ell\Big\}.\label{probS1}
\end{equation}
The solution to \eqref{probS1} is the largest possible value of $m_\ell$ for which at least $U\:t_\ell^\prime$ users experience a TB error probability equal to or smaller than $\Hat{p}$. 

Let $\textrm{P}(\mathbf{N}_{1:\ell})$ be the user-agnostic version of~\eqref{eq:P2} representing the probability of recovering the $\ell$-th expanding window when $\Hat{p}$ is the erasure probability of all TB transmissions. The value of $N_\ell$ is obtained by iteratively solving, for \mbox{$\ell = s+1, \ldots, L$}, the problem below\footnote{Since S1-$\ell$ and S2-$\ell$ are characterized by small discrete feasible spaces, they are simple enough to permit the use of exhaustive search to find their solutions.} (lines~\ref{s2SolStart}-\ref{s2SolEnd})
\begin{equation}
\text{(S2-$\ell$)} \,\, \displaystyle\mathop{\arg\min}_{N_\ell \in [0, \ldots, \Hat{N}_\ell]} \Big\{\textrm{P}(\mathbf{N}_{1:\ell}) \geq \Hat{Q} \,\,\wedge\,\, 0 \leq N_\ell \leq \Hat{N}_\ell\Big\}.\label{probS2}
\end{equation}
where $\wedge$ denotes the logical AND operator.

Both the MCS indexes and the number of TB transmissions associated to each expanding window obtained so far define the \emph{intermediate solution} of the heuristic UEP-RAM $\{m_{\textrm{INT},1}, \ldots, m_{\textrm{INT},L}, N_{\textrm{INT},1}, \ldots, N_{\textrm{INT},L}\}$. If the solution of S2-$\ell$ does not meet constraint~\eqref{UEP.c1}, the value of $s$ is decreased and the while loop continues until all valid values of $s$ are exhausted. If constraint~\eqref{UEP.c1} is fulfilled, a refinement procedure (lines~\ref{refstepStart}-\ref{refstepEnd}) aiming at improving the quality of the intermediate solution is initiated. In particular, the MCS index $m_\ell$ for $\ell \in [s+2, L]$ is potentially decreased by setting it to $m_{\ell-1}$, $N_{\ell-1}$ is set to $0$ and $N_\ell$ is recomputed. We remark that if $N_\ell$ is set to zero, coded packets associated to $\mathbf{x}_{1:\ell}$ will not been generated and transmitted. If the refined solution requires a total number of TB transmissions which is smaller than that associated with the intermediate solution, the procedure returns the refined solution. Otherwise, the procedure returns the intermediate solution.

\section{Numerical Results}\label{sec:Res}
The performance of the proposed resource allocation framework is investigated in this section. We consider a network composed of $19$ base stations, each of which controls three hexagonal cell sectors. In the case of a SC-eMBMS network, $18$ interfering base stations are organized in two concentric rings centred on the base station delivering the target scalable video stream. In the case of a SFN-eMBMS configuration, we consider a SFN formed by $4$ base stations surrounded by the remaining $15$ interfering base stations. Table~\ref{tab.params} describes the main simulation parameters and target values for two H.264/SVC video streams, namely stream A and B~\cite{6025326}.

\begin{table}[tbp]
\vspace{-2.5mm}
\centering
\caption{Simulation parameters and the video streams considered.}
\label{tab.params}
{\scriptsize\begin{tabular}{|c||c|p{29.6mm}|}
\hline \multicolumn{2}{|c|}{\textbf{Paramter}} 	& 	\textbf{Value}						\\
\hline \multicolumn{2}{|c|}{Inter-Site-Distance (ISD)}		& 	$500$ \textrm{m}								\\
\hline \multicolumn{2}{|c|}{System Bandwidth}		& 	$20$ \textrm{MHz}								\\
\hline \multicolumn{2}{|c|}{Duplexing Mode}				& 	FDD									\\
\hline \multicolumn{2}{|c|}{Carrier Frequency}				& 	$2.0$ \textrm{GHz}							\\
\hline \multicolumn{2}{|c|}{Transmission Power}				& 	46 \textrm{dBm} per-sector					\\
\hline \multicolumn{2}{|c|}{Base station and user Antenna Gains}		& 	see Table A.2.1.1-2~\cite{TR_36_814}	\\
\hline \multicolumn{2}{|c|}{Pathloss and Penetration Loss} & 	see Table A.2.1.1.5-1~\cite{TR_36_814}\\
\hline \multicolumn{2}{|c|}{Channel Model} & 	ITU-T PedA~\cite{Access2013}\\
\hline \multicolumn{2}{|c|}{$H$} & 	$2$ KB\\
\hline \multicolumn{2}{|c|}{$n_\ell/N_{\text{RBP}}$,} & 	$2, 3, 5, 6, 8, 10, 12, 14, 17,$\\
\multicolumn{2}{|c|}{for $m_\ell = 4, \ldots, 15$} & 	$20, 66, 72$\\
\hline \multicolumn{2}{|c|}{$\Hat{Q}$} & 	$0.99$\\

\hline \multicolumn{2}{|c|}{$\Hat{N}_\ell$, for $\ell = 1, \ldots, L$,} & 	$\hspace{-1mm}\left\lceil k_\ell/n_{\textrm{min}}\right\rceil \!+\! \left\lceil \Hat{p} \left\lceil k_\ell/n_{\textrm{min}}\right\rceil\right\rceil $,\\

\multicolumn{2}{|c|}{where $n_{\textrm{min}} = \displaystyle\min_{m_\ell \in [4, \ldots, 15]}n_\ell$} & $n_{\textrm{min}} = 2N_{\text{RBP}}$\\

\hline \multirow{1}{*}{Stream A~\cite{StreamA}} & $\{\rho_1, \ldots \rho_3\}$ [dB] & $\hspace{-1.5mm}\{27.9, 35.9, 45.8\}$\\
($L = 3$)& $\{b_1, \ldots, b_3\}$ [Kbps] & $\hspace{-1.5mm}\{47.3, 326.1, 1396.7\}$\\
& $\{\Hat{t}_1, \ldots, \Hat{t}_3\}$ & $\hspace{-1.5mm}\{0.99, 0.8, 0.6\}$\\
\hline \multirow{1}{*}{Stream B~\cite{StreamB}} & $\{\rho_1, \ldots \rho_4\}$ [dB] & $\hspace{-1.5mm}\{28.1, 33.4, 39.9, 46.4\}$\\
($L = 4$) & $\{b_1, \ldots, b_4\}$ [Kbps] & $\hspace{-1.5mm}\{36.8, 79.4, 303.4, 835.9\}$\\
& $\{\Hat{t}_1, \ldots, \Hat{t}_4\}$ & $\hspace{-1.4mm}\{0.99, 0.9, 0.75, 0.6\}$\\
\hline \multirow{1}{*}{Stream A, B} & $d_{\text{GoP}}$& $0.533$ \textrm{s}\\\hline
\end{tabular}}
%\vspace{-2mm}
\end{table}

To assess the quality of the UEP-RAM solutions generated by our proposed heuristic strategy, we shall compare them to resource allocation solutions obtained by directly solving the UEP-RAM using a genetic strategy~\cite{sivanandam2007introduction}, hereafter referred to as \emph{direct solution}. Genetic approaches are known to provide a tight approximation of the optimum solution to the considered problem class~\cite{Deep2009505}. Both the direct and the heuristic solutions shall be compared in terms of the profit-cost ratio $\tau = \sum_{u = 1}^U\sum_{\ell = 1}^L \delta_{u,\ell} \big/ \sum_{\ell = 1}^L N_\ell$ that they can achieve. For clarity, we remark that genetic strategies are characterized by a considerable computational complexity, which makes them unsuitable for use in practical network scenarios~\cite{sivanandam2007introduction}.

To assess the effectiveness of the proposed UEP-RAM, we compare it against the MrT resource allocation idea proposed in~\cite{5452675,4917957}, which relies on a standard LTE-A MAC layer and does not employ NC or AL-FEC error protection strategies. If $\rho_\ell$ denotes the highest Peak Signal-to-Noise Ratio (PSNR) achieved by user $u$ after the recovery of the first $\ell$ layers~\cite{6025326}, while $\dot{\mathrm{P}}_{u,\ell}$ is the probability that user $u$ will recover the first $\ell$ layers, the MrT strategy defines the \emph{user performance level} as
\begin{equation}
\dot{\rho}_u = \displaystyle\max_{\ell = 1, \ldots, L} \left\{\rho_\ell \cdot \dot{\mathrm{P}}_{u,\ell}\right\}.\label{mrt.upl}
\end{equation}
The considered MrT allocation method aims at maximizing the sum of all user performance levels by optimising the MCSs used for the delivery of each layer of the video stream. The MrT optimization problem can be summarized as
\begin{align}
\text{(MrT)} \quad & \max_{m_1, \ldots, m_\ell} \sum_{u=1}^{U} \dot{\rho}_u\\
\text{subject to}   \quad & m_{\ell-1} < m_\ell &  \text{$\ell = 2, \ldots, L$}\label{mrt.c1}
\end{align}
where constraint~\eqref{mrt.c1} allows the model to adopt a different MCS for each video layer and, thus, exploit user heterogeneity in terms of radio propagation conditions.

For the sake of clarity, note that we refer to the approximation of $p_{u,l}$, described in Section~\ref{sec:RA} only during the resource resource allocation operations. When assessing the user QoS, as presented in the following sections, we refer to the actual PDU error probability experienced by each user.

\subsection{SC-eMBMS Delivery Mode}\label{sec.SC}
In order to better study the behaviour of the proposed \mbox{UEP-RAM} strategy, we 
adopt a user disposition pattern that ensures significant user heterogeneity. In the case of \mbox{SC-eMBMS}, users are placed on the radial line representing the symmetry axis of one of the cell sectors, which is controlled by the base station transmitting the target video service. We consider $U = 80$ users, which are spaced $2$ \textrm{m} apart with the first user being placed $90$ \textrm{m} from the base station transmitting the target video service.

\begin{figure}[t]
\centering
\includegraphics[width=1\columnwidth]{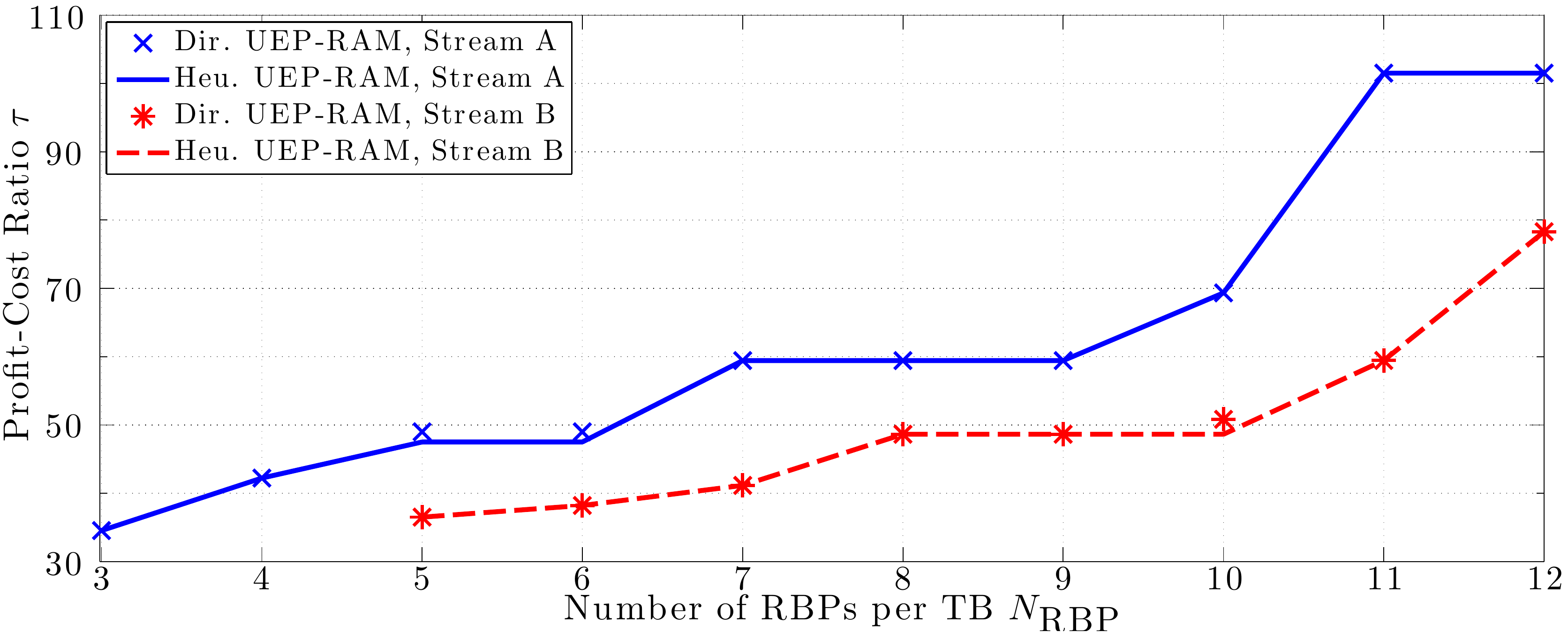}\vspace{-2mm}
\caption{Profit-cost ratio associated with the direct (``Dir.'') and the heuristic (``Heu.'') solution of the UEP-RAM model vs. the number of RBPs per-TB.}
\vspace{-4.5mm}
\label{fig.tau}
\end{figure}

Fig.~\ref{fig.tau} compares the profit-cost ratios associated with resource allocation solutions obtained by directly solving the UEP-RAM and by using the proposed heuristic strategy, as a function of $N_{\text{RBP}}$ for both stream A and stream B. We observe that the performance gap between the direct and the heuristic solutions of UEP-RAM is negligible. Heuristic solutions deviate at most $4.3$\% from direct solutions in Fig.~\ref{fig.tau}. Having established that the proposed heuristic strategy generates good quality solutions, we utilize it in the performance comparison of the MrT and UEP-RAM methods.

\setlength{\textfloatsep}{\oldtextfloatsep}

We first draw attention to the fact that $\Hat{t}_\ell$, which is the fraction of users that shall receive the first $\ell$ video layers with probability of at least $\Hat{Q} = 0.99$, can also be interpreted as a distance from the centre of the cell. For example, if we consider the value $\Hat{t}_4 = 0.6$ and we recall the user disposition pattern that we have adopted, we understand that $0.6\cdot 80=48$ of the users, who are located within a distance of $90+47\cdot 2=184$ \textrm{m} from the centre of the cell, shall recover the first four video layers with probability of at least $\Hat{Q} = 0.99$.

\begin{figure}[t]
\centering
\subfloat[Stream A]{\label{fig.probA}
\includegraphics[width=1\columnwidth]{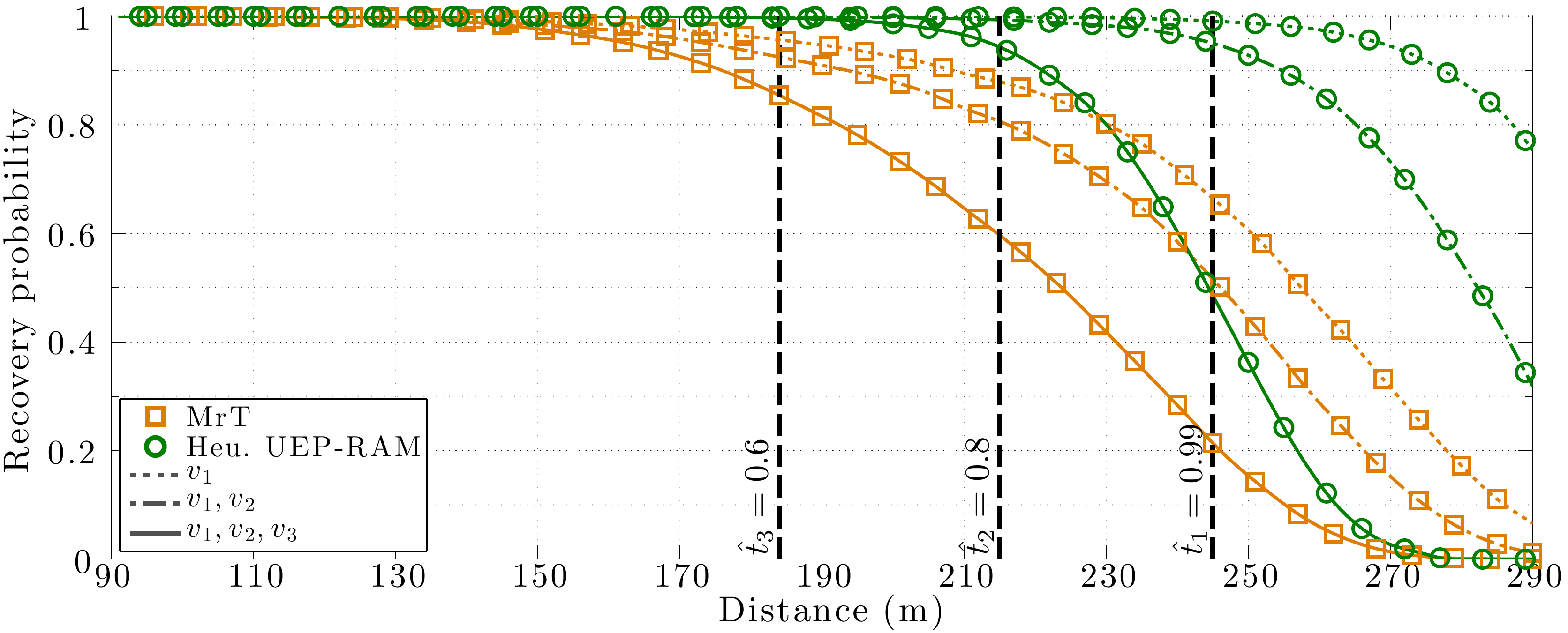}
}\\
\vspace{-2mm}\subfloat[Stream B]{\label{fig.probB}
\includegraphics[width=1\columnwidth]{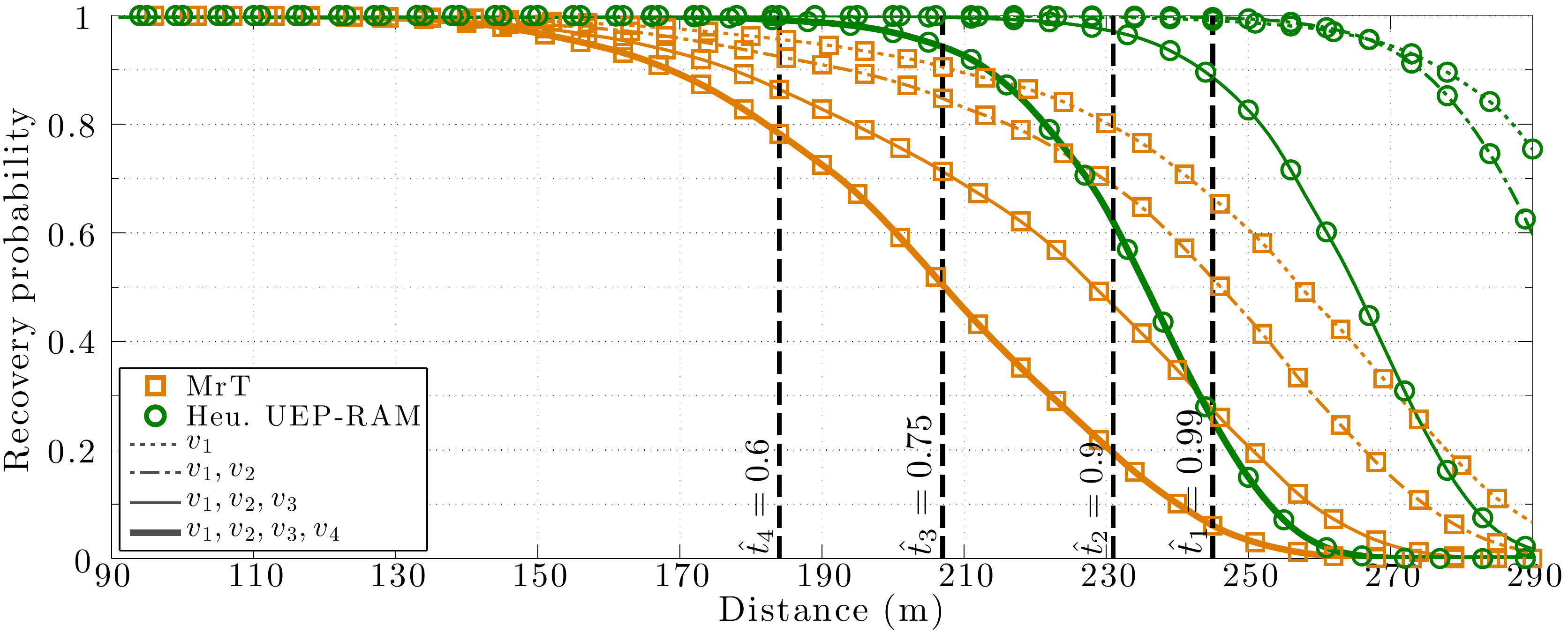}
}
\caption{Probability of recovering the first $\ell$ video layers (with $\ell = 1, \ldots, L$) vs. the distance from centre of the cell.}
\vspace{-4.5mm}
\label{fig.prob}
\end{figure}

Figs.~\ref{fig.probA} and~\ref{fig.probB} show the probability of recovering the first $\ell$ video layers as a function of the distance from the base station in the centre of the cell. Vertical dashed lines denote the target distance for each QoS level. We observe that the proposed UEP-RAM comfortably meets the service coverage constraints. In addition, UEP-RAM ensures a service coverage which is considerably greater than that of MrT. For instance, if UEP-RAM is applied on stream A (Fig.~\ref{fig.probA}), the coverage area will be $64$\% greater than that of MrT for base layer provisioning (i.e., layer $v_1$ only) and $41$\% greater for all-layer provisioning (i.e., layers $v_1$, $v_2$ and $v_3$). The same trend is observed for stream B (Fig.~\ref{fig.probB}).

\subsection{SFN-eMBMS Delivery Mode}\label{sec.SFN}
Similarly to the SC-eMBMS scenario, the user distribution in the SFN-eMBMS case is characterized by significant heterogeneity. We consider a network composed of $U = 1700$ users located at the vertices of a regular square grid with step $20$ \textrm{m}. The user grid is placed on the playground in such a way that the area spanned by the SFN is uniformly covered, starting from inside the SFN. 

In this section, we use the maximum PSNR experienced by a user as the performance metric. As in the previous section, the adopted user distribution allows us to convert the maximum PSNR at each user into contours of the playground. Consequently, a visual comparison of the coverage areas offered by MrT and UEP-RAM can be performed. We explained that the maximum PSNR for the MrT strategy, denoted by $\dot{\rho}_u$, can be obtained from \eqref{mrt.upl}. The formulation of the maximum PSNR for the proposed UEP-RAM can be easily adapted from \eqref{mrt.upl} as follows
\begin{equation}
\rho_u = \displaystyle\max_{\ell = 1, \ldots, L} \left\{\rho_\ell \cdot \mathrm{P}_u(\mathbf{N}_{1:\ell})\right\}.\label{uep-ram.upl}
\end{equation}

\begin{figure}[t]
\centering
\subfloat[Stream A]{\label{fig.sfnA}
\hspace{-1.6mm}\includegraphics[width=0.5\columnwidth]{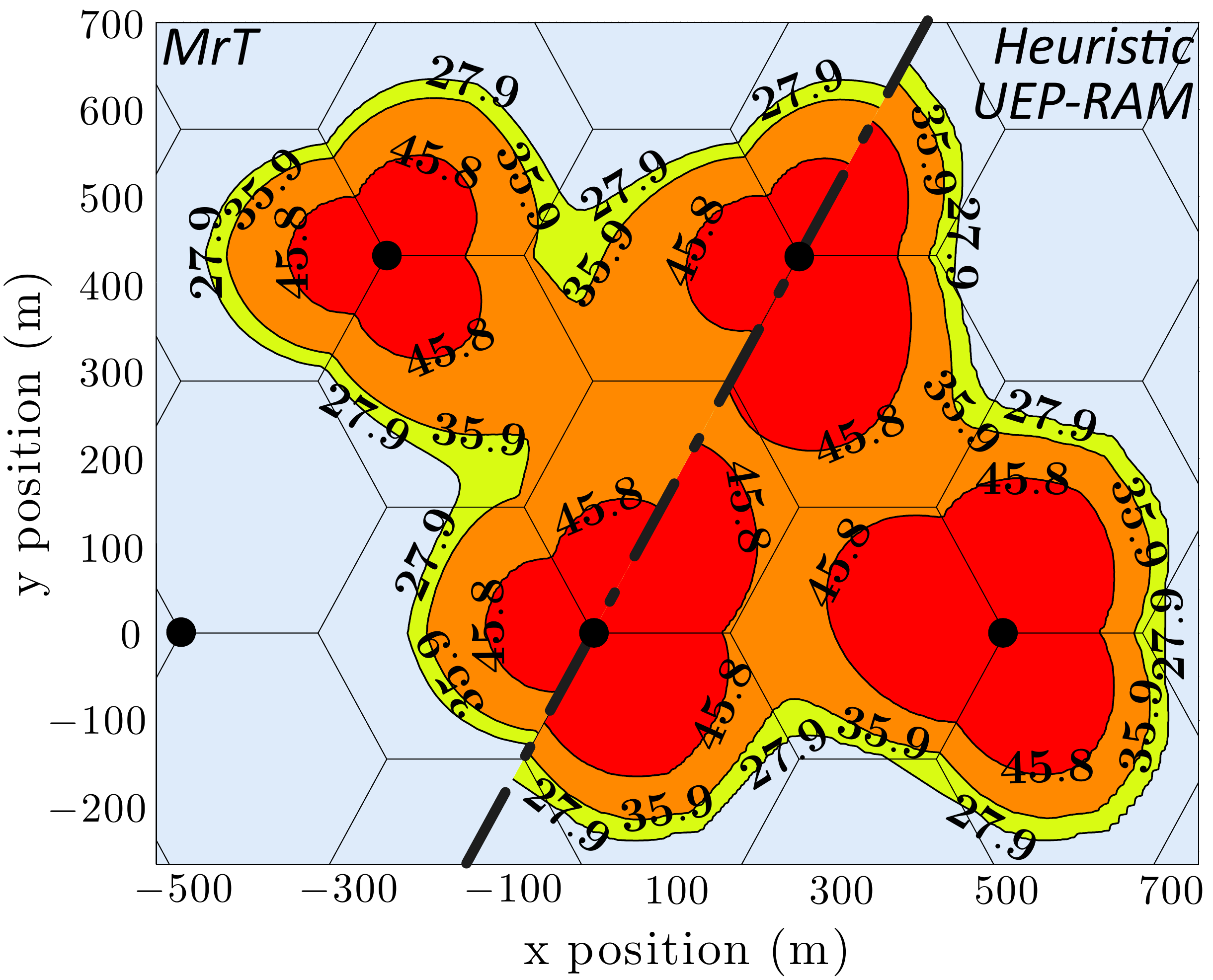}
}
\subfloat[Stream B]{\label{fig.sfnB}
\hspace{-1.6mm}\includegraphics[width=0.5\columnwidth]{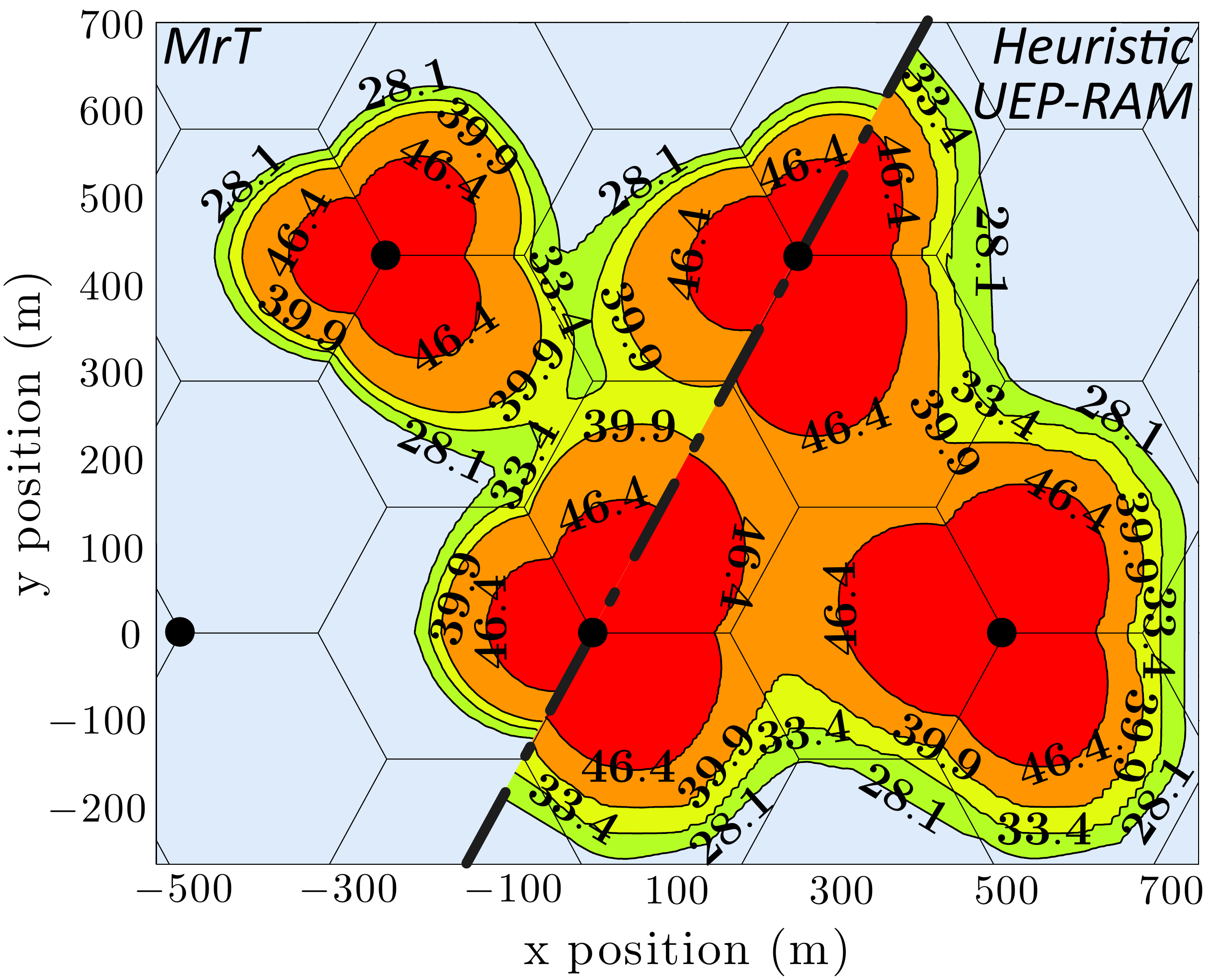}
}
\caption{Maximum PSNR values (dB) associated with the MrT and heuristic UEP-RAM strategies (black circles represents the base stations).}
\vspace{-4mm}
\label{fig.sfn}
\end{figure}

Fig.~\ref{fig.sfn} shows the aforementioned performance metric associated with each point of the playground for the two resource allocation strategies and the two video streams when $N_{\text{RBP}} = 5$ RBPs. Coloured regions depict areas where users can achieve a maximum PSNR value that is equal to or greater than that of $\rho_\ell$ shown in Table~\ref{tab.params}. Table~\ref{tab.sfnProb} reports the fraction of users in Figs.~\ref{fig.sfnA} and~\ref{fig.sfnB} that can recover the first $\ell$ video layers, for $\ell = 1, \ldots, L$, with at least a target probability $\Hat{Q}=0.99$. We observe that the heuristic UEP-RAM ensures a service coverage which is considerably greater than that of MrT. For instance, the heuristic UEP-RAM resource allocation solutions allow $65$\% of users to recover all four video layers of stream B and all users to recover the base layer. In comparison, the MrT strategy successfully delivers all four video layers to about $33$\% of users, while the base layer is received by less than $40$\% of users.

\section{Conclusions}\label{sec:Concl}
In this paper we presented a viable method for the incorporation of the UEP-NC scheme into LTE-A networks as a means of improving the reliability of a H.264/SVC video multicast service. Furthermore, we proposed a novel resource allocation framework that aims to improve the service coverage with a reduced resource footprint. Inspired by a fundamental economics principle, the goal of the proposed optimization framework is achieved by maximizing the system profit-cost ratio. In spite of the natural complexity of the proposed modelling, we defined a novel heuristic strategy that was used to efficiently derive a good quality resource allocation solution of the considered problem. Numerical results showed that the proposed resource allocation framework ensures a service coverage which is up to $2.5$-times greater than that of the considered conventional MrT strategy.

\section*{Acknowledgemt}
This work is part of the R2D2 project, which is supported by EPSRC under Grant EP/L006251/1. Collaboration of the authors was facilitated by COST Action IC1104 on Random Network Coding and Designs over $\mathrm{GF}(q)$.

\bibliographystyle{IEEEtran}
\bibliography{biblio}

\begin{table}[t]
\centering
\caption{Fraction of users recovering each set of video layers.}
\label{tab.sfnProb}
{\scriptsize
\begin{tabular}{|c||c|c||c|c|} 
\hline 
\multirow{3}{*}{\hspace{-1mm}Recovered Layers\hspace{-1mm}} & \multicolumn{2}{c||}{Stream A} & \multicolumn{2}{c|}{Stream B}\\
 \cline{2-5}
 & \multirow{2}{*}{MrT} & Heur. & \multirow{2}{*}{MrT} & Heur.\\
 &  & UEP-RAM &  & UEP-RAM \\
  \hline
$v_1$  & $39.9$\% & $100$\% & $39.9$\% & $100$\%\\
  \hline
$v_1, v_2$  & $36.1$\% & $81$\% & $36.1$\% & $100$\%\\
  \hline
$v_1, v_2, v_3$  & $34.1$\% & $74$\% & $34.3$\% & $87.4$\%\\
  \hline
$v_1, v_2, v_3, v_4$  & - & - & $33.1$\% & $65$\%\\
  \hline
\end{tabular}
}
\end{table}
\end{document}